\documentclass[aps,pre,floats,twocolumn,superscriptaddress,showpacs]{revtex4}

\usepackage{epsfig}
\usepackage{latexsym,amsmath}

\newcommand{\avk}{\langle k\rangle}
\newcommand{\fluck}{\langle k^2\rangle}

\begin{document}

\title{Generation of uncorrelated random scale-free networks}

\author{Michele Catanzaro}
\affiliation{Departament de F\'\i sica i Enginyeria Nuclear, Universitat
  Polit\`ecnica de Catalunya, Campus Nord B4, 08034 Barcelona, Spain}

\author{Mari\'an Bogu\~n\'a}
\affiliation{Departament de F\'\i sica Fonamental, Universitat de
  Barcelona, Av. Diagonal 647, 08028 Barcelona, Spain}

\author{Romualdo Pastor-Satorras}
\affiliation{Departament de F\'\i sica i Enginyeria Nuclear, Universitat
  Polit\`ecnica de Catalunya, Campus Nord B4, 08034 Barcelona, Spain}

\date{\today}

\begin{abstract}
  Uncorrelated random scale-free networks are useful null models to
  check the accuracy an the analytical solutions of dynamical
  processes defined on complex networks. We propose and analyze a
  model capable to generate random uncorrelated scale-free networks
  with no multiple and self-connections. The model is based on the
  classical configuration model, with an additional restriction on the
  maximum possible degree of the vertices. We check numerically that
  the proposed model indeed generates scale-free networks with no two
  and three vertex correlations, as measured by the average degree of
  the nearest neighbors and the clustering coefficient of the vertices
  of degree $k$, respectively.
\end{abstract}


\maketitle

Complex networks constitute a general framework for the topological
characterization of many natural and technological systems whose
complexity prevents a more detailed microscopic description
\cite{barabasi02,dorogorev,mendesbook}. Within this framework, these
systems are represented in terms of networks or graphs
\cite{bollobas98}, in which vertices stand for the units composing the
system, while edges among vertices represent the interactions or
relations between pairs of units. The focus is thus shift to the
topological characterization of the representative network, a task
which is largely more feasible and yields, nevertheless, a noticeable
amount of information on the structure and properties of the original
system.  The empirical analysis of many real complex networks has
unveiled the presence of several typical properties, widely found in
systems belonging to a large variety of realms. One of the most
relevant is given by the scale-free nature of the degree distribution
$P(k)$ \cite{barabasi02,mendesbook,barab99}, defined as the
probability that a randomly chosen vertex has degree $k$ (i.e. it is
connected to other $k$ vertices). In mathematical terms, the
scale-free property translates into a power-law function of the form
\begin{equation}
  P(k) \sim k^{-\gamma},
  \label{eq:1}
\end{equation}
where $\gamma$ is a characteristic degree exponent. The presence of a
scale-free degree distribution can have an important impact on the
behavior of dynamical processes taking place on top of the network.
Indeed, scale-free networks with exponent $\gamma$ in the range $2 < \gamma \leq 3$
show large fluctuations in their degrees, evident in the presence of a
diverging second moment, $\langle k^2 \rangle$, in the infinite network size limit
$N \to \infty$. This divergency, in turn, show up in a remarkable weakness of
the network in front of targeted attacks \cite{havlin01,newman00} or
the propagation of infectious agents \cite{pv01a,lloydsir}.

It has been recently realized that, besides their degree distribution,
real networks are also characterized by the presence of degree
correlations. This translates in the observation that the degrees at
the end points of any given edge are not usually independent. This
kind of degree-degree correlations can be theoretically expressed in
terms of the conditional probability $P(k'|k)$ that a vertex of degree
$k$ is connected to a vertex of degree $k'$. From a numerical point of
view, it is more convenient to characterize degree-degree correlations by
means of the average degree of the nearest neighbors of the vertices
of degree $k$, which is formally defined as \cite{alexei}
\begin{equation}
  \bar{k}_{nn}(k) = \sum_{k'} k' P(k'|k).
  \label{knn}
\end{equation}
Degree-degree correlations has led to a first
classification of complex networks according to this property
\cite{assortative}. Thus, when $\bar{k}_{nn}(k)$ is an increasing
function of $k$, the corresponding network is said to exhibit
\textit{assortative mixing by degree}, i.e. highly connected vertices are
preferentially connected to highly connected vertices and vice-versa,
while a decreasing $\bar{k}_{nn}(k)$ function is typical of
\textit{disassortative mixing}, being highly connected vertices more
probably connected to poorly connected ones. For uncorrelated networks, the
degrees at the end points of any edge are completely independent. Therefore,
the conditional probability $P(k'|k)$ can be simply estimated as the
probability that any edge points to a vertex of degree $k'$, leading
to $P_{\mathrm{nc}}(k'|k) = k' P(k')/ \avk$, independent of $k$. Inserting this equation in Eq. (\ref{knn}), the average nearest neighbors degree reads
\begin{equation}
\bar{k}_{nn}^\mathrm{nc}(k) = \frac{\fluck}{\avk}, \label{eq:3}
\end{equation}
that is, independent of the degree $k$.

Analogously, from a theoretical point of view, correlations concerning three vertices can be
characterized by means of the conditional probability $P(k'', k'|k)$
that a vertex of degree $k$ is simultaneously connected to two
vertices of degrees $k'$ and $k''$. We can estimate this
kind of three point correlations by means of the clustering coefficient of the
vertices of degree $k$, $\bar{c}(k)$ \cite{alexei02,ravasz02}, defined
as the probability that two neighbors of a vertex of degree $k$ are
also neighbors themselves. This function can be formally written as
\begin{equation}
  \bar{c}(k) = \sum_{k', k''} P(k'', k'|k) p_{k', k''},
  \label{eq:2}
\end{equation}
where $p_{k', k''}$ is the probability that vertices $k'$ and $k''$
are connected given that they have a common neighbor
\cite{hiddenvars}.  For random uncorrelated networks, the three vertex
conditional probability can be written as $P_{\mathrm{nc}}(k'', k'|k)
= P_{\mathrm{nc}}(k''|k) P_{\mathrm{nc}}(k'|k)$, for $k>1$. The
connection probability can also be computed in this case as $p_{k',
  k''}=(k'-1)(k''-1)/\langle k \rangle N$, where the term $-1$ comes from the fact
that one of the connections of each vertex has already been used
\cite{hiddenvars,newmanrev,dorogovtsevclustering}. From the above
relations, the clustering coefficient becomes
\begin{equation}
   \bar{c}_\mathrm{nc}(k) = \frac{(\fluck - \avk)^2}{\avk^3 N}. \label{eq:4}
\end{equation}
As in the previous case, for uncorrelated networks, the function
$\bar{c}(k)$ is constant and independent of $k$. Therefore, any non
trivial dependence of the functions $\bar{k}_{nn}(k)$ and $\bar{c}(k)$
on the degree is a signature of the presence of two and three point
correlations respectively.

While most real networks show indeed the presence of correlations,
uncorrelated random networks are nevertheless equally important from a
practical point of view, especially as null network models in which to
test the behavior of dynamical systems whose analytic solution is
usually available only in the absence of correlations
\cite{havlin01,newman00,pv01a,moreno02}. Therefore,
it becomes an interesting issue the possibility to generate random
networks which have a guaranteed lack of correlations. In the
particular case of scale-free networks, however, finding such
algorithms is far more difficult than one would expect {\it a priori}.
In this paper, we observe that classical algorithms, which are
supposed to generate uncorrelated networks, do, indeed, generate
correlations when the desired degree distribution is scale-free and no
more than one edge is allowed between any two vertices
\cite{maslovcorr,originnewman}. To solve this drawback, we present and
test an algorithm capable to generate uncorrelated scale-free
networks.

The classical algorithm to construct random networks with any
prescribed degree distribution $P(k)$ is the so-called configuration
model (CM)
\cite{bekessi72,benderoriginal,molloy95,molloy98,newmanrev}.  To
construct a network with the original definition of this algorithm, we
start assigning to each vertex $i$, in a set of $N$ vertices, a random
degree $k_i$ drawn from the probability distribution $P(k)$, with $m \leq
k_i \leq N$ (no vertex can have a degree larger than $N$), and imposing
the constraint that the sum $\sum_i k_i$ must be even. The network is
completed by randomly connecting the vertices with $\sum_i k_i /2$ edges,
respecting the preassigned degrees.  The result of this construction
is a random network whose degrees are, by definition, distributed
according to $P(k)$ and in which, in principle, there are no degree
correlations, given the random nature of the edge assignment.

While this prescription works well for bounded degree distributions,
in which $\fluck$ is finite, one has to be more careful when dealing
with networks with a scale-free distribution, which, for $2< \gamma \leq 3$,
yield diverging fluctuations, $\fluck\to\infty$, in the infinite network size
limit.  In fact, it is easy to see that, if the second moment of the
degree distribution diverges, a completely random assignment of edges
leads to the construction of an uncorrelated network, but in which a
non-negligible fraction of self-connections (a vertex joined to
itself) and multiple connections (two vertices connected by more than
one edge) are present \cite{mariancutofss}.  While multiple and
self-connections are completely natural in mathematical graph theory
\cite{bollobas98}, they are somewhat undesired for simulation
purposes, since most real network do not display such structures, and
also in order to avoid ambiguities in the definition of the network
and any dynamics on top of it. This situation can be avoided by
imposing the additional constraint of forbidding multiple and
self-connections. This constraint, however, has the negative side
effect of introducing correlations in the network
\cite{maslovcorr,originnewman}.
\begin{figure}
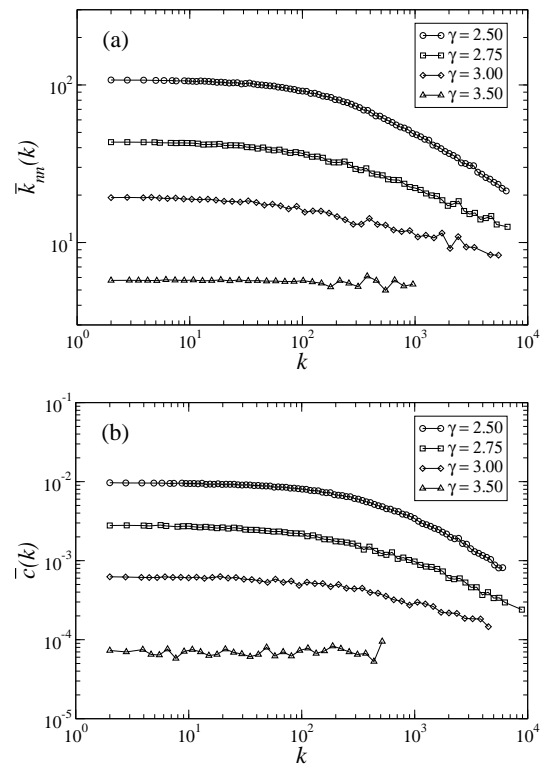

  \epsfig{file=fig1a.eps, width=7cm}\vspace*{0.25cm}
  \epsfig{file=fig1b.eps, width=7cm}
  \caption{Average nearest neighbor degree of the vertices of degree
    $k$, $\bar{k}_{nn}(k)$ (a) and average clustering coefficient
    $\bar{c}(k)$ (b) for the original CM algorithm with different degree
    exponents $\gamma$. Network size is $N=10^5$.}
  \label{fig:cm}
\end{figure}
As an example of this fact, in Fig.~\ref{fig:cm} we show the functions
$\bar{k}_{nn}(k)$ and $\bar{c}(k)$ computed from numerical simulations
of the CM algorithm with no multiple and self-connections for
different $\gamma$ exponents and fixed network size $N=10^5$. As we can
observe, for $\gamma>3$, which corresponds to an effectively bounded degree
distribution with finite $\fluck$, both functions are almost flat,
signaling an evident lack of correlations. On the other hand, for
values $\gamma\leq 3$ there is a clear presence of correlations. This
correlations have a mixed disassortative nature: vertices with many
connections tend to be connected to vertices with few connections,
while low degree vertices connect equally with vertices of any degree.

The origin of this phenomenon can be traced back to the effects of the
cut-off (or maximum expected degree) $k_c(N)$ in a network of size
$N$.  In fact, it is possible to show that in order to have no
correlations in the absence of multiple and self-connections, a
scale-free network with degree distribution $P(k) \sim k^{-\gamma}$ and size
$N$ must have a cut-off scaling at most as $k_s(N) \sim N^{1/2}$ (the
so-called structural cut-off) \cite{mariancutofss}.  For a power-law
network generated using the CM algorithm defined above (i.e.
generating random degrees in the range $m \leq k_i \leq N$), simple extreme
value theory arguments show in fact that
\begin{equation}
  k_c(N) \sim N^{1/(\gamma-1)}.
  \label{eq:19}
\end{equation}
For $\gamma<3$, we have that $k_c(N) > N^{1/2}$ and therefore it is
impossible to avoid the presence of correlations. Only for the
particular case $\gamma\geq 3$ we recover $k_c(N) \leq  N^{1/2}$, which explains
the lack of correlations observed in Fig.~\ref{fig:cm} for $\gamma=3.5$.

\begin{figure}
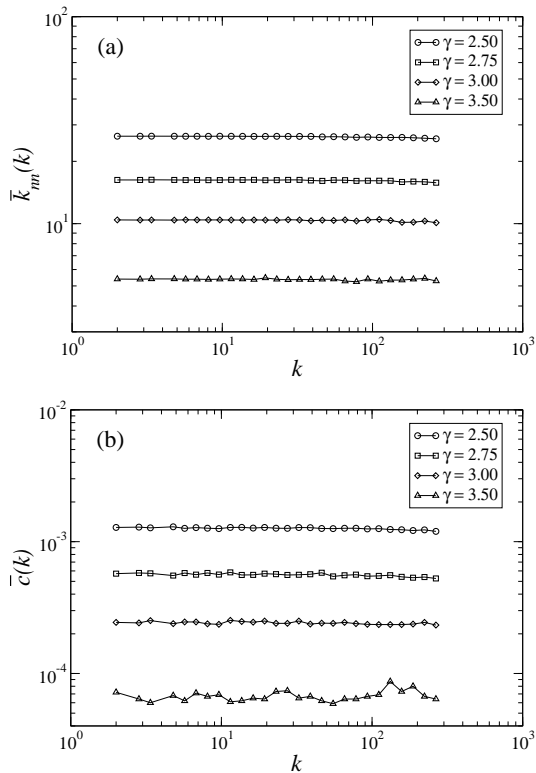

  \epsfig{file=fig2a.eps, width=7cm}\vspace*{0.25cm}
  \epsfig{file=fig2b.eps, width=7cm}
  \caption{Average nearest neighbor degree of the vertices of degree
    $k$, $\bar{k}_{nn}(k)$ (a) and average clustering coefficient
    $\bar{c}(k)$ (b) for the UCM algorithm with different degree
    exponents $\gamma$. Network size is $N=10^5$.}
  \label{fig:ucm}
\end{figure}

Since it is the maximum possible value of the degrees in the network
that rules the presence or absence of correlations in a random network
with no multiple or self-connections, we propose the following
\textit{uncorrelated configuration model} (UCM) in order to generate
random uncorrelated scale-free networks:
\begin{enumerate}
\item Assign to each vertex $i$ in a set of $N$ initially disconnected
  vertices a degree $k_i$, extracted from the probability distribution
  $P(k)\sim k^{-\gamma}$, and subject to the constraints $m \leq k_i \leq N^{1/2}$
  and $\sum_i k_i$ even.
 \item Construct the network by randomly connecting the vertices with
   $\sum_i k_i /2$ edges, respecting the preassigned degrees and avoiding
   multiple and self-connections.
\end{enumerate}
The constraint on the maximum possible degree of the vertices ensures
that $k_c(N) \sim N^{1/2}$, allowing for the possibility to construct
uncorrelated networks.  As an additional numerical optimization of
this algorithm, we also impose the minimum degree $m=2$ to generate
connected networks with probability one \cite{molloy98,havlin00}.

In Fig.~\ref{fig:ucm} we check for the presence of correlations in the
UCM model for scale-free networks. As we can
observe, both correlation functions show an almost flat behavior for
all values of the degree exponent $\gamma$, compatible with the lack of
correlations at the two and three vertex levels.

\begin{figure}
  \epsfig{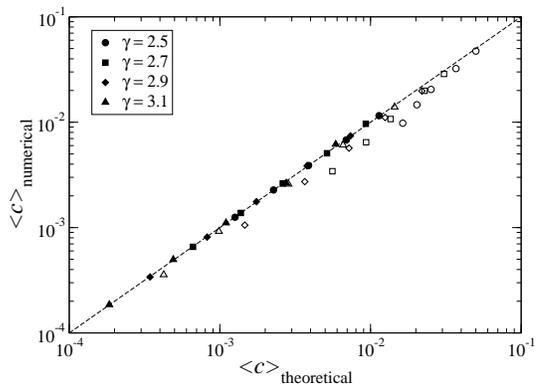}
  \caption{Numerical average clustering coefficient $\langle c \rangle$ as a
    function of the corresponding theoretical value, given by
    Eq.~(\ref{eq:4}), for the CM (hollow symbols) and the UCM (full
    symbols) algorithms. The different points for each value of $\gamma$
    represent different network sizes $N=10^3$, $3 \times 10^3$, $10^4$, $3
    \times 10^4$, and $10^5$.}
  \label{fig:clustering}
\end{figure}

We have additionally explored the validity of the expression for the
average clustering coefficient \cite{watts98}, $\langle c \rangle$, defined as
\begin{equation}
  \langle c \rangle = \sum_k P(k) \bar{c}(k),
\end{equation}
which, for random uncorrelated networks takes the form given by
Eq.~(\ref{eq:4}) \cite{newmanrev}. For scale-free networks with a
general cut-off $k_c(N)$, we have that, in the large $N$ limit,
$\fluck \sim k_c(N)^{3-\gamma}$. Therefore, for random networks generated with
the classical CM model, in which $k_c(N) \sim N^{1/(\gamma-1)}$, we have that
$\langle c \rangle_\mathrm{CM} \sim N^{(7-3\gamma)/(\gamma-1)}$. This expression is clearly
anomalous for $\gamma < 7/3$, since it leads to a diverging clustering
coefficient for large $N$, while, by definition, this magnitude,
being a probability, must be smaller that one. This anomaly vanishes
in the UCM prescription. In this case, we have that $k_c(N) \sim N^{1/2}$
for any value of $\gamma$, leading to $\langle c \rangle_\mathrm{UCM} \sim N^{2-\gamma}$, which
is a decreasing function of the network size for any $\gamma> 2$. 

In Fig.~\ref{fig:clustering} we plot the average clustering
coefficient obtained from numerical simulations of the CM and UCM
algorithms as a function of the theoretical value, Eq. (\ref{eq:4}),
for different values of $\gamma$ and different network sizes $N$. We can
observe that, while the results for the uncorrelated UCM model nicely
collapse into the diagonal line in the plot, meaning that the
numerical values are almost equal to their theoretical counterparts,
noticeable departures are observed for the implicitly correlated CM
algorithm.

To sum up, in this paper we have presented a model to generate
uncorrelated random networks with no multiple and self-connections and
arbitrary degree distribution. The lack of correlations is especially
relevant for the case of scale-free networks. In this case, our
algorithm is capable to generate networks with flat correlation
functions and an average clustering coefficient in good agreement with
theoretical predictions. Our algorithm is potentially interesting in
order to check the accuracy of the many analytical solutions of
dynamical processes taking place on top of complex networks, which are
usually found in the uncorrelated limit and, which, up to now, lacked
a proper benchmark to check the results for degree exponents $\gamma < 3$.

\vspace*{0.5cm}

This work has been partially supported by EC-FET Open Project No.
IST-2001-33555.  R.P.-S. acknowledges financial support from the
Ministerio de Ciencia y Tecnolog\'\i a (Spain), and from the Departament
d'Universitats, Recerca i Societat de la Informaci\'o, Generalitat de
Catalunya (Spain). M. B. acknowledges financial support from the
Ministerio de Ciencia y Tecnolog\'\i a through the Ram\'on y Cajal program.
M. C. acknowledges financial support from Universitat Polit\`ecnica de
Catalunya.

\end{document}